\documentclass{elsart5p}

 \usepackage{graphicx}

\usepackage{amssymb}

\usepackage{lineno}

\begin{document}

\begin{frontmatter}



\title{Measurement of acoustic properties of South Pole ice for neutrino astronomy}


\author{Justin Vandenbroucke}
\ead{justinav@berkeley.edu}
\author{for the}
\author{IceCube Collaboration}
\ead[url]{http://www.icecube.wisc.edu}
\address{Department of Physics, University of California, Berkeley, CA 94720, USA}



\begin{abstract}
South Pole ice is predicted to be the best medium for acoustic neutrino detection.  Moreover, ice is the only medium in which all three dense-medium detection methods (optical, radio, and acoustic) can be used to monitor the same interaction volume.  Events detected in coincidence between two methods allow significant background rejection confidence, which is necessary to study rare GZK neutrinos.  In 2007 and 2008 the South Pole Acoustic Test Setup (SPATS) was installed as a research and development project associated with the IceCube experiment.  The purpose of SPATS is to measure the acoustic ice properties at the South Pole in order to determine the feasibility of a future large hybrid array.  The deployment and performance of SPATS are described, as are first results and work in progress on the sound speed, background noise, and attenuation.
\end{abstract}

\end{frontmatter}

\begin{figure*}
\noindent\includegraphics[width=40pc]{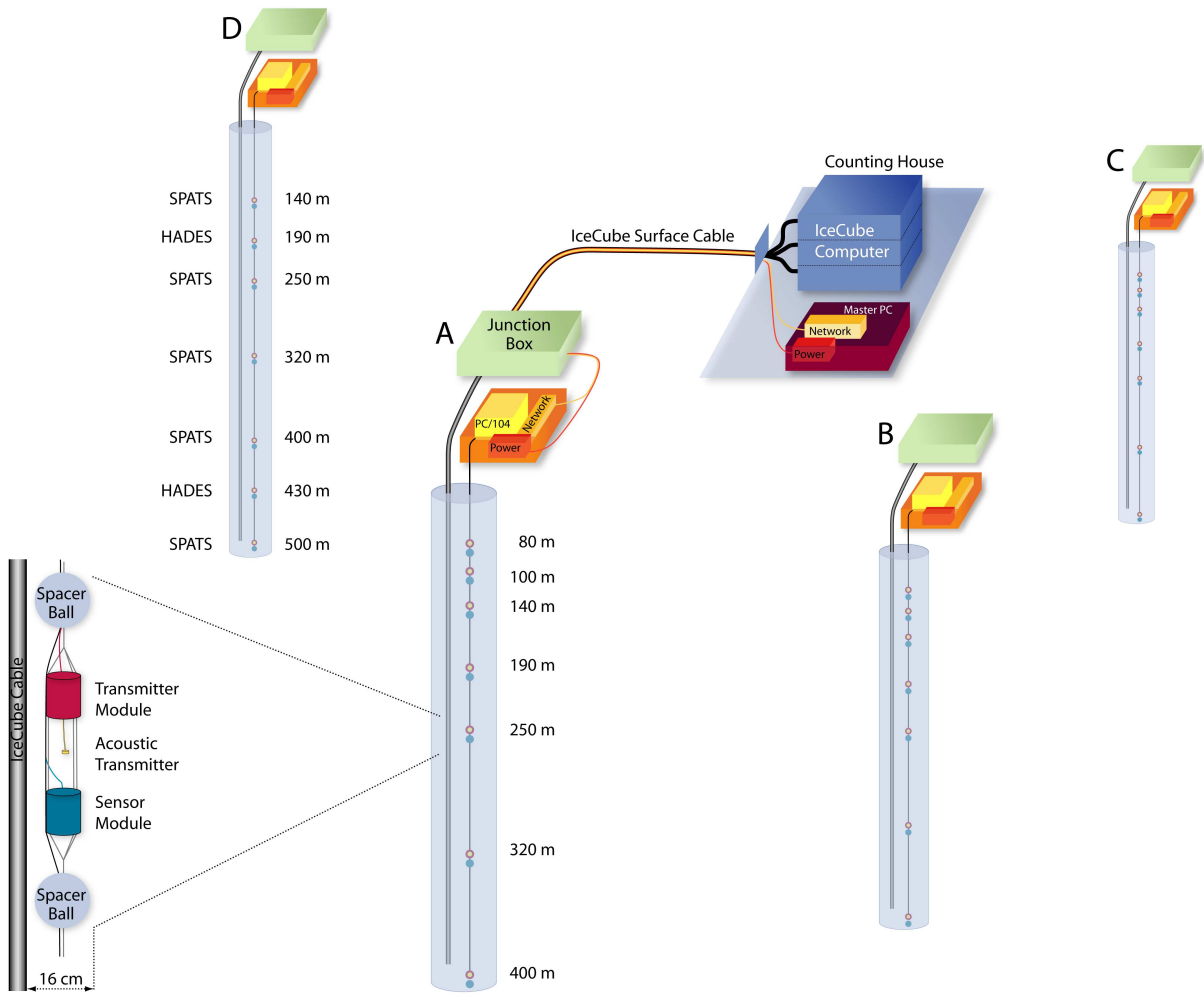}
\caption{Schematic of the SPATS array.}
\label{spats_schematic_4_strings}
\end{figure*}

\section{Motivation}

Along with photons and charged cosmic rays, neutrinos are the third leg of the tripod necessary to understand high energy astro-particle physics.  In particular, detecting and characterizing the cosmogenic or ``GZK'' neutrinos produced by ultra-high-energy charged particles interacting with the cosmic microwave background would contribute essential understanding to solve the problem of ultra-high-energy particle acceleration.  Charged particle paths are scrambled by magnetic fields, and at the highest energies we can only detect cosmic rays from sources within $\sim$1 cosmic ray interaction length (``GZK length'', $\sim$50 Mpc).  The propagation length of high-energy photons is similarly limited by their interaction with background photon fields.

Cosmic rays from sources at cosmological distances interact before reaching us.  However, the interaction produces GZK neutrinos that we can detect.  Moreover, GZK neutrinos point within $\sim$1 GZK length of the source and therefore the extremely high energy cosmic ray sources can be resolved with neutrinos if they are at cosmological distances.  In fact, due to the small scattering angle in this interaction, the neutrinos point to the cosmic ray source within an even smaller radius than the GZK length.  Unlike photons and charged cosmic rays, neutrinos only interact with background fields if their energy is above $\sim$10$^{21}$~eV~\cite{Weiler82}.

In addition to building a GZK neutrino sky map, measuring the spectrum of the GZK neutrinos would help constrain ultra-high-energy cosmic ray production and propagation.  Furthermore, the zenith distribution of detected events could be used to measure the total neutrino-nucleon cross section at $\sim$100 TeV center-of-mass energy~\cite{2006aren.conf..163C}.  This fundamental measurement would test models of quantum gravity and extra dimensions, and cannot be made with any envisioned Earthbound particle accelerator.  In addition to astrophysics and particle physics, neutrinos with energy above 10$^{18}$~eV may also be valuable for cosmology~\cite{Ringwald:2006ks}.

In the next few years, ANITA, IceCube, and the Pierre Auger Observatory will either discover GZK neutrinos or determine that their flux is at the low end of the range of theoretical predictions.  If the flux is sufficiently large that GZK neutrino astronomy is feasible, the next step after discovery will be to characterize the signal using angular, temporal, and spectral distributions, for the reasons described above.  A follow-up experiment will require enough sensitivity to detect on the order of 100 events in a few years.  This experiment would benefit from improved signal reconstruction and background rejection capabilities, compared to the currently operating experiments for which discovery is the priority.  One promising approach is to detect a subset of events with more than one method simultaneously, using a hybrid detector.  South Pole ice is uniquely suited for hybrid neutrino detection, with the possibility to detect high energy neutrino interactions in the ice using the optical, radio, and acoustic signals they generate.  The AMANDA and IceCube experiments have set stringent limits on the high energy neutrino flux by searching for their optical signals~\cite{Ackermann:2007km}.  The RICE experiment has set limits by searching for neutrino-induced radio signals~\cite{Kravchenko06}.

In~\cite{Besson05}, a $\sim$100~km$^3$ hybrid extension of IceCube featuring optical, radio, and acoustic neutrino detection was simulated.  The array components were exposed to the same Monte Carlo neutrino set, and neutrinos detected by any one method or any combination of them were identified.  The simulated array detected more than 10 GZK neutrinos per year.  Moreover, while only a few percent were detected in coincidence between the optical and radio method or the optical and acoustic method, $\sim$40$\%$ were detected in coincidence between the radio and acoustic methods.

Although a large array is necessary to achieve a significant event rate, an intermediate array built on a shorter time scale could provide essential technical experience in the design, construction, and operation of a hybrid array.  Radio and acoustic sensors spaced at large distances will have power and communications requirements that are similar to each other and different from IceCube, and a prototype to test such infrastructure would be valuable.  A possible intermediate array was described in~\cite{Besson08}.

To determine the feasibility of an acoustic GZK neutrino detector, the acoustic properties of the South Pole ice need to be measured.  The South Pole Acoustic Test Setup (SPATS) was designed, in association with the IceCube experiment, to do this.  Its goal is to determine, \emph{in situ}, the acoustic noise floor, sound speed as a function of depth, rate and nature of background transients, and attenuation length, in the frequency range relevant to neutrino detection (1-100~kHz).

The attenuation length of acoustic waves in South Pole ice is predicted to depend on temperature (and therefore depth).  The predicted value is several km in the shallowest, coldest 1-2~km of ice~\cite{Price06}.  In order to verify that a large acoustic neutrino experiment could detect a significant rate of cosmogenic neutrinos, the attenuation length should be established to be on the order of 1~km.  An attenuation length of several km is not necessary, because for values above $\sim$1~km attenuation is dominated by $1/r$ divergence.

A description of the initial SPATS deployment and first results was given in~\cite{Boeser08}.

\section{Instrumentation}

\subsection{Array}

\begin{figure}
\noindent\includegraphics[width=20pc]{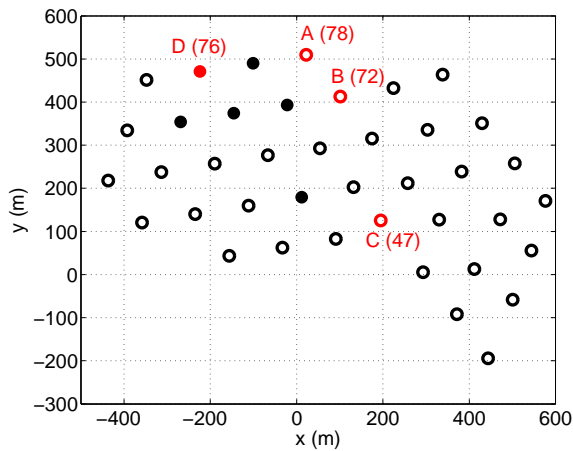}
\caption{Surface layout of the SPATS array.  All 40 IceCube strings deployed as of February 2008 are shown.  The 4 holes that contain both SPATS and IceCube strings are labeled.  The holes in which the retrievable pinger was operated are indicated with filled circles.}
\label{geometry_2008}
\end{figure}

A schematic of the SPATS array is given in Figure~\ref{spats_schematic_4_strings}.  The array consists of four strings, each deployed in an IceCube hole alongside an IceCube string (Figure~\ref{geometry_2008}).  Each string has seven acoustic stages, with each stage consisting of one transmitter module and one sensor module.  Strings A, B, and C were deployed in January 2007 and contain stages at depth 80, 100, 140, 190, 250, 320, and 400~m.  String D was deployed in December 2007 and contains stages at depth 140, 190, 250, 320, 400, 430, and 500~m.  The horizontal baselines are AB = 125~m, AC = 421~m, AD = 249~m, BC = 302~m, BD = 330~m, and CD = 543~m.

Each transmitter module consists of a ring-shaped piezoelectric transducer molded in epoxy and in direct contact with the ice, connected to a pressure housing containing electronics to pulse the transducer.  Each sensor module contains three disk-shaped piezoelectric transducers inside a single cylindrical pressure housing.  The three channels point in different azimuthal directions, each separated by 120$^\circ$ from its neighbors, to obtain good angular coverage.

After deployment of the first three strings, a fourth string (String D) was determined necessary to complete our attenuation measurement.  Results from the initial data were used to improve this string relative to the first three.  The shallowest stages were eliminated and replaced with deeper stages.  String D contains second-generation sensors and transmitters of a design similar to those in the first three strings, as well as an alternative (``HADES'') sensor design featuring a complementary dynamic range and different impedance matching to the ice~\cite{Semburg08}.

The stages of each string are connected to the surface via a copper cable assembly.  Buried beneath $\sim$2~m of ice at the surface of each string is a junction box receiving this cable.  A rugged embedded computer (``String PC'') in the junction box controls the transmitters on the string and digitizes the analog waveforms received from the sensors on the string.  The String PC at the surface of each string communicates via an Ethernet extender with a central, indoor computer (``Master PC'').  A GPS signal of the IRIG-B format is distributed to each of the String PC's, where it is digitized synchronously with sensor and transmitter data to achieve $\sim$10~${\mu}$s absolute time stamping precision.

\subsection{Retrievable pinger}

In December 2007 - January 2008, a retrievable pinger was operated in several IceCube holes.  The pinger consists of a spherical piezoelectric emitter\footnote{International Transducer Company ITC-1001} connected to a pressure housing containing a high voltage pulser circuit.  The pinger was connected to the surface via an armored, four-conductor cable which transmitted the pinger power and trigger signal.  The pinger was triggered with a one-pulse-per-second signal from a GPS clock, such that the absolute emission time of each pinger pulse was known (modulo one second).  The pinger was lowered to a maximum depth of 400-500~m, then raised back to the surface, while it pinged continuously at 1 Hz.

At each of the depths with SPATS stages, the pinger lowering was stopped for five minutes.  Throughout each pinger deployment, all SPATS strings were recording their sensor channels.  On each string the data acquisition (DAQ) software looped over the 21 sensor channels (17 on String D), recording the channels one at a time due to the limited bandwidth between each String PC and the Master PC.  Each sensor recording was nine seconds in duration and sampled at 200~kHz.  The three/four (before/after String D deployment) strings recorded simultaneously with one another.  The five minute stop at each level ensured that every sensor channel of the array recorded the pinger emitting nine consecutive pulses while it was stopped at each depth.

\subsection{Performance}

Each string was connected to its junction box and commissioned within $\sim$24~hours of deployment.  Commissioning proceeded smoothly, with all strings receiving power, communications, and timing from the Master PC successfully.  22 months (11 for String D) after deployment, SPATS continues to perform well.  Despite being buried two meters under the ice surface at -50$^\circ$~C, the String PC's are running smoothly.  There have been several unexpected power outages that suddenly cut power to the String PC's and the in-ice instrumentation, in some cases lasting up to $\sim$48~hours.  In each case the String PC's and in-ice strings have restarted without incident.  

93$\%$ of sensor channels are behaving normally.  A few channels are dead or continuously oscillating in saturation (presumably due to a lost ground contact).  The number of healthy sensor channels is 18 of 21 on String A, 21 of 21 on String B, 18 of 21 on String C, and 17 of 17 on String D.

The String PC ADC clocks, used to drive sampling of sensor waveforms as well as to pulse the transmitters at a fixed repetition rate, drift at a rate on the order of a few ${\mu}$s/s.  That is, the true sampling frequency differs from the nominal frequency by an amount on the order of a few parts per million.  Many of our analyses use a recording of multiple pulses generated by a source repeating at a fixed repetition rate, and we wrap the waveforms in time and average them to improve our signal-to-noise ratio.  If the nominal ADC sampling frequency is used for this instead of the true one, the pulse averaging decoheres.  We measure the true sampling frequency of each clock as needed with the IRIG-B GPS signal.

\subsection{Calibration}

The sensors were calibrated in laboratory water tanks prior to deployment.  The main purpose of this calibration was to estimate the angular variation of the sensor module response, as well as the module-to-module variation.  \emph{In situ} calibration may be possible, in particular to determine the relative module to module variation with a single calibration source such as the retrievable pinger.

The absolute response of each sensor (\emph{i.e.}, the conversion from Pascals to Volts) once it is in the ice is different from that in the laboratory, due to the different temperature, pressure, and medium (solid ice compared with liquid water).  Work is underway to determine the variation in sensor response due to variation of each of these parameters.  In air, the amplitude of a signal recorded from a transmitter was found to increase linearly as the temperature of the sensor was decreased: The signal increased by $\sim$40$\%$ when the sensor was cooled from 0$^\circ$~C to -50$^\circ$~C.  After the ARENA conference where these results were presented, we measured the variation of sensor response with pressure, in a liquid-filled laboratory pressure vessel.  We found the sensor response variation between 0 and 100~bar to be less than 20$\%$.  100~bar corresponds to a hydrostatic pressure of 1~km water equivalent, twice the maximum depth at which we have deployed sensors.  Finally, a laboratory setup featuring a large tank in a freezer room has been constructed to compare the response of the sensors in water and ice.

\section{Results}

\subsection{Pressure waves and shear waves}

With both the frozen-in transmitters and the retrievable pinger, we have detected both pressure and shear wave pulses with the SPATS sensors.  The shear waves from the frozen-in transmitters are likely produced at the emitter itself, as expected if there is a shear component to the ring-shaped piezoelectric element's response to the applied electrical pulse.  The retrievable pinger, on the other hand, was operated in a water-filled IceCube drill hole of diameter $\sim$60~cm, before other instrumentation was deployed in it.  Because shear waves cannot propagate through fluids, the shear waves detected from the retrievable pinger were likely produced when the pressure waves encountered the water/ice interface (the hole wall).  This ``mode conversion'' from pressure to shear waves is favored under glancing angle of incidence, and disfavored under normal incidence.

The problem of pressure wave transmission, pressure wave reflection, and mode conversion to produce shear wave transmission at a liquid-to-solid interface is a complex one, but it must be considered in detail in order to estimate the attenuation length of the ice using the pinger data.  The pressure transmission coefficient, pressure reflection coefficient, and shear transmission coefficient all vary with the angle of incidence (within the constraint of energy conservation).  If the sensor and transmitter are at different depths, or if the pinger is not within the plane determined by the central axis of its hole and the sensor in a different hole, the incident angle is non-zero and shear wave production is expected.

The amplitude of the shear pulses relative to the pressure pulses varies considerably from pulse to pulse in the retrievable pinger data.  It is not surprising that the partitioning between pressure and shear amplitudes varies from pulse to pulse and run to run, given that the pinger was free to move laterally (pendulum mode) and vertically (bouncing mode) within the water column, and that the transmission coefficients vary with incident angle.

The hypothesis that pulse-to-pulse and run-to-run variation in the retrievable pinger waveforms is due to motion of the source is supported by the data.  Sensor waveforms recorded while the pinger was being lowered or raised exhibit significant pulse-to-pulse variation, as do some runs recorded soon after the pinger was stopped at a particular depth.  Waveforms recorded several minutes after the pinger was stopped at any given depth generally exhibit less pulse-to-pulse variation.  Furthermore, the shear to pressure ratio often varied from pulse to pulse within a single nine-second run, indicating that the angle of incidence varied over the same short time span.  Finally, the shear and pressure amplitudes are anti-correlated, as expected under the constraint of energy conservation.

In both pressure and shear pulses, secondary pulses are often detected which are separated from the primary pulse by a time up to $\sim$0.8~ms.  This time corresponds to the expected separation between the direct signal and the reflection off the ``back'' wall of the pinger hole as seen from the sensor, for pinger-to-back-wall distance up to $\sim$0.6~m.  This time varies from run to run and pulse to pulse, further evidence that there is pinger motion which complicates the data.  In some cases the direct and reflected signal are well separated; in other cases the two mostly overlap and interfere significantly.

\subsection{Sound speed}

\begin{figure}
\noindent\includegraphics[width=20pc]{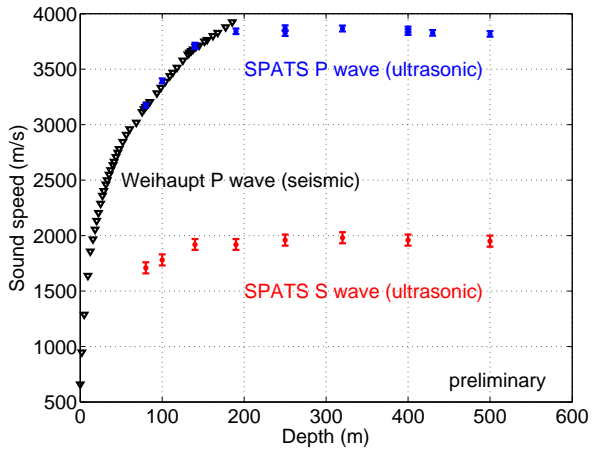}
\caption{Sound speed versus depth, as measured with SPATS, for both pressure and shear waves.  $\pm$1$\sigma$ error bars are indicated.  The results from Weihaupt~\cite{Weihaupt63}, for which no uncertainty estimate was given, are shown for comparison.}
\label{spats_and_weihaupt}
\end{figure}

We used pinger data to measure the speed of both pressure waves and shear waves as a function of depth with better than 3$\%$ precision (Figure~\ref{spats_and_weihaupt}).  The shear wave speed is $\sim1/2$ the pressure wave speed, as expected~\cite{martinb:ice-density}.  In the firn (the shallowest $\sim$200~m, where the ice is not fully compactified), our pressure wave speed results agree with a previous measurement~\cite{Weihaupt63}.  We have completed the first measurement of pressure wave speed beneath the firn and of shear wave speed in both the firn and bulk (fully compactified) ice.  Both the pressure wave speed and shear wave speed increase with depth in the firn and are independent of depth in the bulk ice.

In the fully compactified ice, we measure the pressure wave speed to be 3850 $\pm$ 50~m/s, and the shear wave speed to be 1950 $\pm$50~m/s.  Moreover, we find that the speed gradient for both pressure and shear waves is consistent with zero.  Future improvements in the precision of this measurement will improve the upper limit on the gradient, or will resolve it if it is small but non-zero.  The sound speed measurement is described in detail in~\cite{Descamps08}.

Precise knowledge of the sound speed is a requirement for reconstructing transient events, including neutrino signals, and separating them from backgrounds.  Refraction increases the complexity of event reconstruction algorithms and decreases their precision.  If the amount of refraction is small, reconstruction is straightforward, fast, and precise.  With sufficiently small refraction it is possible to use fully analytical event reconstruction algorithms.

\subsection{Noise floor}

The distribution of sample amplitudes at any particular sensor channel is described well by a normal distribution.  On each string, the noise level decreases with increasing depth.  This is consistent with the expectation that the sound speed gradient in the shallowest $\sim$200~m refracts surface noise back to the surface, shielding it from the deep ice.  The noise level at each sensor channel is stable over time, featuring no correlation with surface conditions.  There is no correlation with season, temperature, or wind, and no correlation with the intense human surface activity involving heavy machinery in the summer season.  This is further evidence that refraction is shielding the deep ice from both natural and anthropogenic surface noise, and is encouraging for neutrino detection.

The noise on each sensor has been observed to increase during drilling of each IceCube hole by the IceCube hot water drill.  In fact, SPATS sensors heard the drilling of each of the 18 holes drilled in December 2007 - January 2008, including those greater than 600~m distant.

Due to the challenge of \emph{in situ} absolute calibration, it is difficult to convert the noise level at each channel from Volts to Pascals, which is necessary to estimate the neutrino energy threshold achievable with the acoustic technique.  Nevertheless, an estimate is possible if several assumptions are made in extrapolating laboratory results to \emph{in situ} conditions.  Under these assumptions, we estimate the total noise (Gaussian $\sigma$, integrating the noise between 10 and 50~kHz) in the deep ice to be $<$~10~mPa.  A neutrino with energy on the order of 10$^{18}$~eV, interacting at 1~km distance, is expected to produce a signal of amplitude $\sim$10~mPa at a point in the middle of the disk-shaped radiation pattern.  The noise results obtained with SPATS are presented in detail in~\cite{Karg08}.

\subsection{Attenuation}

We have performed two classes of attenuation analysis: inter-string (using frozen-in transmitters) and pinger (using the retrievable pinger operated in water).

In the first year of operation, we used inter-string data from the first three strings (with frozen-in transmitters) to estimate the attenuation length~\cite{Boeser08}.  Our first analysis used the most straightforward approach: combining all transmitter-sensor paths, including those that are diagonal (have sensor and transmitter at different depth).  No attempt was made to correct for angular variation in the sensors and transmitters.  However, the transmitters and sensors are known to have zenith angular variation in their response, and in particular to be most responsive in the horizontal plane.  Therefore the unaccounted angular response not only added scatter to the data, but is expected to result in a systematic underestimation of the attenuation length, because the diagonal paths are generally longer than the horizontal ones.

\begin{table}
\centering
\caption{Systematic effects present in various attenuation analyses, and their influence on the attenuation result.  Each effect tends to increase the attenuation length estimate, decrease the attenuation length estimate, or have a varying effect on it (depending on the data set used).} 
\begin{tabular}{| c | c |}  
\hline                        
\bf{Systematic effect} & \bf{Effect on $\lambda_{att}$ estimate} \\
\hline                    
Interference with hole wall reflections & varying \\
\hline
Polar angular variation in sensitivity & decreases \\
\hline
Residual clock drift & varying \\
\hline
Saturation (\emph{i.e.}, limited dynamic range) & increases \\
\hline
Hole ice & varying \\
\hline
Noise, if not subtracted & increases \\
\hline
Water-ice transmission coefficient & decreases \\
\hline
Channel to channel variation & varying \\
\hline     
\end{tabular} 
\label{systematic_effects}  
\end{table} 

Because we do not have sufficient angular calibration data for each module to account for this effect, we have focused on using transmitters and sensors confined to a single depth.  However, with only three strings all transmitters at any particular depth needed to be combined and then the analysis was sensitive to transmitter-to-transmitter variation.  This is one of the reasons that we added String D.

The motivation behind the retrievable pinger was (1) to use a single calibration source recorded simultaneously by different sensors while it is in the same location, and (2) to record with each single sensor the same calibration source when the source is in different locations.  We expected this to reduce the systematic effects caused by module to module variation in the three-string inter-string analyses.

We have applied several different analyses to the pinger data.  One analysis considered a single pinger deployment recorded by a single sensor, using the amplitude as a function of distance as the pinger was lowered down and up in its hole.  This has the benefit of using only a single transmitter and a single receiver.  However, it suffers from systematic effects due to the angular response of the sensor (for which no correction was attempted), and to the variation of the water-to-ice transmission coefficient with angle.

Other analyses exploited a valuable feature of the pinger hole geometry: Three pinger deployments were in holes along a straight line intersecting String B.  This means that individual sensor channels could be selected from String B, and their amplitude plotted versus pinger distance.  This has the advantage of eliminating sensor-to-sensor variation as well as azimuthal response variation of individual sensors.  One particular analysis used this geometry applied to the three sensor channels at depth 320~m.  We calculated the signal energy of each pulse, summing the contributions from pressure and shear waves using time windows determined from the known propagation speeds, and subtracting the noise energy estimated from off-time regions of the waveform.  The pressure and shear energies were summed because of the observed pulse-to-pulse variation in both the pressure and shear amplitudes, and the anti-correlation between them.

Other analyses used amplitudes calculated in the frequency domain.  This has increased power in separating signal from noise and is less sensitive to pulse-to-pulse variations in waveform shape in the time domain.  Peak-to-peak and ``first peak'' amplitudes were also used.  The conclusion from the pinger analyses is that the existing data are too complex, with too many systematic effects, to derive a robust attenuation estimate.  Systematic effects present in both pinger and inter-string analyses are summarized in Table~\ref{systematic_effects}.

\begin{figure}
\noindent\includegraphics[width=20pc]{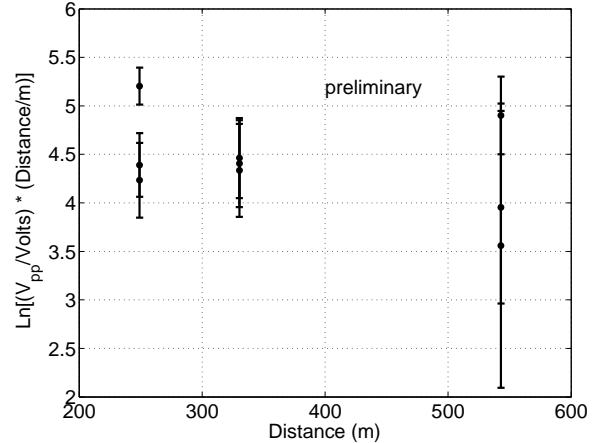}
\caption{Natural log of (amplitude times distance) versus distance, for nine sensor channels recording a single frozen-in transmitter, with the source and all receivers at 320~m depth.}
\label{amp_vs_dist_log}
\end{figure}

Now that String D has been deployed, it is possible to pulse one frozen-in transmitter and record three different sensor modules, each with three channels, at the same depth as the transmitter.  This significantly reduces the number of variables and therefore the number of systematic effects.  Because only one transmitter is used, we do not need to make any assumptions about transmitter-to-transmitter variation.  Because only one depth is used, zenith variation of the transmitter and sensors is not relevant.  Furthermore we use a frozen-in transmitter, which eliminates the pulse-to-pulse variation due to pinger motion as well as the water/ice interface effects.

Shortly after the ARENA conference, this analysis strategy was applied at the 320~m depth, with the String D transmitter pulsing and the three sensor channels at the same depth on each of the three other strings recording.  Figure~\ref{amp_vs_dist_log} shows the peak-to-peak amplitude recorded at each channel as a function of distance.  400 consecutive pulse waveforms were averaged, accounting for clock drift, to determine each data point.  Error bars indicate the statistical error of the pulse averaging algorithm.

\begin{figure}
\noindent\includegraphics[width=20pc]{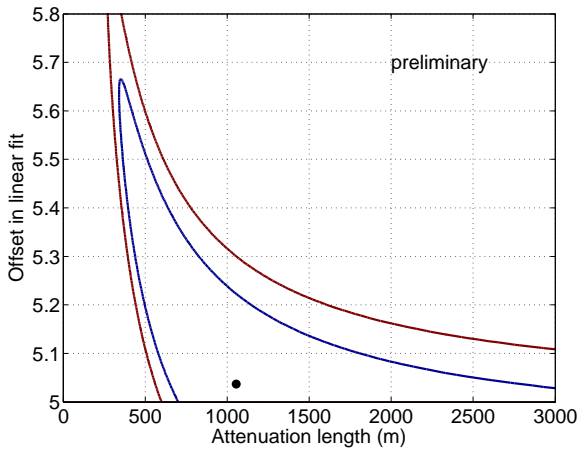}
\caption{Confidence region for the acoustic attenuation length (and the offset in the linear fit), for the single-transmitter, single-depth four-string analysis.  The best fit is shown along with inclusion contours at 68$\%$ and 90$\%$ confidence level.  Systematic errors are not included, and the contours were determined without constraining the attenuation length to be positive: There is another region of allowed parameter space for negative attenuation length which is not visible in this plot.}
\label{attenuation_length_contours}
\end{figure}

A linear fit was applied to these data.  Confidence regions for the fit are given in Figure~\ref{attenuation_length_contours}.  The best fit attenuation length is 1100~m.  The lower limit at 68$\%$ confidence level is 360~m.  The fit and confidence regions were determined assuming zero systematic error and without enforcing any constraint that the attenuation length is positive.

The most significant systematic effect in this analysis is sensor channel-to-channel variation in the absolute response.  The systematic uncertainty from this effect is estimated from water tank calibration data to be $\sim$100$\%$.  Adding this systematic uncertainty in quadrature with the statistical error on the pulse amplitudes, and constraining the attenuation length to be positive, the lower limit is 270~m at 68$\%$ confidence and 170~m at 90$\%$ confidence.  Work is underway to repeat the analysis at other depths, to use pulse energy instead of peak-to-peak amplitude as a more robust amplitude metric, and to apply other analysis techniques to the data.  We are also implementing a new algorithm to improve online clock drift determination, which we will use to generate a new set of inter-string data from which we can calculate the average waveform for each transmitter-sensor combination with improved confidence.

\subsection{Transients}

At the time of the ARENA conference, a DAQ upgrade was underway for transient data taking.  In August 2008, this upgrade was completed.  SPATS runs of 45-minute duration are now being taken every hour, sampling three channels per string simultaneously at 200~kHz.  This marks the beginning of a new phase of data taking: SPATS is now operating as a four-string, twelve-channel transients detector.

We operate with a bipolar discriminator trigger: If the number of ADC counts on any channel exceeds 5.2~$\sigma$ above or below the mean (DC) amplitude, that channel triggers.  When a channel triggers, we record a 5~ms window centered on the triggering sample.  Each channel operates independently, with no online coincidence requirement.  Each channel's DC offset and noise $\sigma$ is used to set a high and low threshold independently from the other channels.  As discussed above, these quantities are stable over time.

This transients DAQ configuration results in a trigger rate of a few events per minute per channel.  This rate is stable over time and roughly equal from channel to channel.  It is slightly larger than the rate expected from Gaussian statistics.  This indicates both that most events are Gaussian noise events where only one sample of the waveform meets the trigger condition, and that there is a fraction of events in excess of the Gaussian noise events.

\begin{figure}
\noindent\includegraphics[width=20pc]{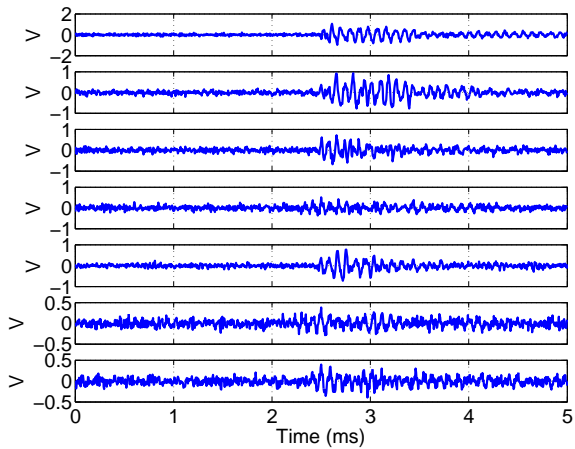}
\caption{Waveforms from an example transient event.  The pulses arrived at the seven different channels within 200~ms of one another.  The trigger of each event occurs at 2.5 ms, the center of the capture window (absolute times of the waveforms are not shown).}
\label{hour01_cluster534_waveforms}
\end{figure}

Offline time-coincidence clustering and source reconstruction algorithms are under development to analyze the transient events.  An example event that triggered seven sensor channels within 200~ms of one another is shown in Figure~\ref{hour01_cluster534_waveforms}.  Preliminary analysis of the transient events indicates good source localization capabilities and verifies that the full chain of online software, offline software, and precision timing is working as expected.  Results will be presented in future publications including~\cite{Vandenbroucke09}.

The transients phase of data taking with SPATS is just beginning.  With further analysis and more data we hope to identify and characterize natural as well as anthropogenic transient sources.  The upgraded DAQ software will operate during the next IceCube drilling season, where we expect it to collect a valuable set of events from the IceCube drill.  Moreover, natural sources such as cracking in the bulk ice or stick-slip motion of the glacier along bedrock could have interesting glaciological implications.  Both anthropogenic and naturally occurring sources could be used for \emph{in situ} calibration of relative sensor response and/or attenuation measurement, by building up large statistics and over-constraining either of the two problems.  All types of transients will also be studied to determine if they constitute a significant background to neutrino detection.

\section{Conclusion and outlook}

SPATS has completed some of its measurement goals, and has made progress on the remaining ones.  The sound speed profile has been measured for both pressure and shear waves.  This has glaciological applications and will be essential input for a future acoustic neutrino detector.  We have determined the noise floor to be stable and Gaussian, to decrease with depth, and to be insensitive to surface noise from both natural and human sources.  The Gaussianity and stability of the noise would make the design and operation of an online trigger in an acoustic neutrino detector straightforward.  This is particularly true in comparison with acoustic detectors deployed in water, where noise conditions vary significantly on multiple timescales, making sophisticated adaptive triggers necessary and making the interpretation of the data challenging~\cite{Vandenbroucke05}.

We have made significant progress on the remaining measurement goals: absolute noise floor, attenuation length, and background transients.  The absolute noise floor has been estimated by extrapolating laboratory calibrations.  Ideas to improve this with \emph{in situ} calibration are under consideration.  Multiple different data types and analysis techniques have been used to estimate the acoustic attenuation length.  Our initial three-string analyses were insufficient to significantly constrain the attenuation.  Many analysis techniques were applied to the retrievable pinger data.  While we derived attenuation constraints from the pinger data, each analysis is subject to different systematic effects and we do not believe that any one of the results from the existing pinger data is sufficiently robust.

We have completed an initial analysis of four-string data using a single frozen-in transmitter to minimize systematic effects resulting from transmiter-to-transmitter variation and from source motion and hole ice interface effects in the retrievable pinger data.  While the uncertainty in this result is large, the attenuation estimate is consistent with theoretical estimates.  Further analysis of these data, using the same technique at different depths as well as other techniques, is underway.

We have improved the pinger design and plan to deploy it at distances of $\sim$125 to $\sim$1000 m from SPATS strings in December 2008 - January 2009.  We have added a mechanical support to the pinger, intended to minimize both the pendulum and bounce modes and to keep it close to the center of the hole.  Furthermore we are increasing the repetition rate of the pinger by more than an order of magnitude, in order to improve the signal to noise ratio after pulse averaging.  We expect the improved pinger, operated at larger baselines, along with new inter-string data and analysis with frozen-in transmitters, to significantly improve our constraints on the attenuation length.

SPATS is now operating as a 4-string, 12-sensor transient event detector.  Analysis of the transient data is ongoing, and first results for determining event sources are promising.  Anthropogenic and natural transients, as well as the upgraded pinger data taking planned for the next season and improved four-string inter-string analysis underway, promise to improve our understanding of the ice and help us determine the feasibility of an acoustic neutrino array at the South Pole.  If such an acoustic array is feasible, it could be combined with a radio array such that a significant fraction of detected neutrinos would be detected in coincidence between the two methods.



\section{Acknowledgments}
The author gratefully acknowledges the support of the National Science Foundation Graduate Research Fellowship Program.

\bibliography{vandenbroucke_arena08}

\begin{thebibliography}{10}
\expandafter\ifx\csname url\endcsname\relax
  \def\url#1{\texttt{#1}}\fi
\expandafter\ifx\csname urlprefix\endcsname\relax\def\urlprefix{URL }\fi

\bibitem{Weiler82}
T.~Weiler, Resonant absorption of cosmic-ray neutrinos by the relic-neutrino
  background, Phys. Rev. Lett. 49~(3) (1982) 234--237.

\bibitem{2006aren.conf..163C}
A.~{Connolly}, {Measuring the Neutrino-Nucleon Cross Section with SalSA}, in:
  R.~{Nahnhauer}, S.~{B{\"o}ser} (Eds.), Acoustic and Radio EeV Neutrino
  Detection Activities, 2006, pp. 163--167.

\bibitem{Ringwald:2006ks}
A.~Ringwald, L.~Schrempp, {Probing neutrino dark energy with extremely
  high-energy cosmic neutrinos}, JCAP 0610 (2006) 012.

\bibitem{Ackermann:2007km}
M.~Ackermann, et~al., {Search for Ultra High-Energy Neutrinos with AMANDA-II},
  Astrophys. J. 675 (2008) 1014.

\bibitem{Kravchenko06}
I.~Kravchenko, et~al., {RICE} limits on the diffuse ultra-high energy neutrino
  flux, Phys. Rev. D73 (2006) 082002.

\bibitem{Besson05}
D.~Besson, S.~B{\"{o}}ser, R.~Nahnhauer, P.~B. Price, J.~Vandenbroucke,
  Simulation of a hybrid optical/radio/acoustic extension to {I}ce{C}ube for
  {E}e{V} neutrino detection, Int. J. Mod. Phys. A21S1 (2006) 259--264.

\bibitem{Besson08}
D.~Besson, R.~Nahnhauer, P.~B. Price, D.~Tosi, J.~Vandenbroucke, B.~Voigt,
  Simulation of a hybrid optical-radio-acoustic neutrino detector at the
  {S}outh {P}ole, These proceedings.

\bibitem{Price06}
P.~B. Price, Attenuation of acoustic waves in glacial ice and salt domes, J.
  Geophys. Res. 111 (2006) B02201.

\bibitem{Boeser08}
S.~B{\"{o}}ser, et~al., {Feasibility of acoustic neutrino detection in ice:
  Design and performance of the South Pole Acoustic Test Setup (SPATS)},
  Proceedings of the 30th ICRC.

\bibitem{Semburg08}
B.~Semburg, {HADES}: {H}ydrophone for {A}coustic {D}etection at the {S}outh
  {P}ole, These proceedings.

\bibitem{martinb:ice-density}
D.~G. Albert, Theoretical modeling of seismic noise propagation in firn at the
  {S}outh {P}ole, {A}ntarctica, Geophysical Research Letters 25~(23) (1998)
  4257--4260.

\bibitem{Weihaupt63}
J.~G. Weihaupt, Seismic and gravity studies at the {S}outh {P}ole, Geophysics
  28~(4) (1963) 582--592.

\bibitem{Descamps08}
F.~Descamps, Measurement of sound speed versus depth with the {S}outh {P}ole
  {A}coustic {T}est {S}etup, These proceedings.

\bibitem{Karg08}
T.~Karg, Acoustic noise in deep ice and environmental conditions at the {S}outh
  {P}ole, These proceedings.

\bibitem{Vandenbroucke09}
J.~Vandenbroucke, PhD thesis, in preparation.

\bibitem{Vandenbroucke05}
J.~Vandenbroucke, G.~Gratta, N.~Lehtinen, Experimental study of acoustic
  ultra-high-energy neutrino detection, Astrophys. J. 621 (2005) 301--312.

\end{thebibliography}
\bibliographystyle{elsart-num}







\end{document}